\documentclass{PoS}

\title{Layered Phase Investigations}

\ShortTitle{Layered Phase Investigations}

\author{\speaker{Stam Nicolis}\\
  Universit\'e ``Fran\c{c}ois Rabelais'' de Tours\\
                 Laboratoire de Math\'ematiques et Physique Th\'eorique
\thanks{CNRS UMR 6083 and F\'ed\'eration Denis Poisson (FR 2964)}\\
Parc Grandmont, 37200 Tours, France
\\
        E-mail: \email{stam.nicolis@lmpt.univ-tours.fr}}

\abstract{The extra dimensional 
defects that are introduced to generate the lattice chiral zero
          modes are not simply a computational trick, but have 
	  interesting physical consequences. After reviewing what is known
	  about the layered phase they can generate, I argue  how it is
	  possible to simulate Yang-Mills theories with reduced systematic
	  errors and speculate on how it might be possible to study the 
	  fluctuations of the layers' topological charge.
          }

\FullConference{The XXV International Symposium on Lattice Field Theory\\
		 July 30 - August 4 2007\\
		 Regensburg, Germany}

\begin{document}

The layered phase was discovered in 1984\cite{fu_nielsen} as a possible way
for dimensional reduction. The authors found that an anisotropic 
 4+1-dimensional lattice gauge theory, with compact $U(1)$ gauge group with
 action (and standard notation)
\begin{equation}
\label{action_gauge}
S_\mathrm{gauge} = \beta\sum_{x}\sum_{\mu<\nu}(1-\mathrm{Re}\left(U_{\mu\nu}(x)\right)) + 
\beta'\sum_{x}\sum_\mu(1-\mathrm{Re}\left(U_{\mu5}(x)\right)) 
\end{equation}
 possessed, in addition to the bulk confining ($\beta$ and $\beta'$ small)
 and Coulomb ($\beta'$  and $\beta$ large) phases, a
 ``layered'' phase, where the Wilson loops within a four-dimensional layer
 followed a perimeter law, whereas those along the extra dimension followed an 
area law.  Of particular interest was that this new phase survived, when
taking the one loop corrections into account, only for four-dimensional
layers, but not for lower-dimensional ones. This invites speculation that it
will subsist beyond the validity  of the (classical) equations of motion. 

A year later, within a totally different context, Callan and
Harvey\cite{callan_harvey}  noticed
that fermions coupled to domain walls in $2n+1$ dimensions have zero modes
localized on the domain wall, whose chirality depends on the sign of the
gradient of the effective mass of the fermion along the extra dimension. Some
years later Kaplan\cite{kaplan} proposed to use this mechanism on the lattice
to evade the Nielsen-Ninomiya theorem and thus define lattice fermions with
exact chiral symmetry--the {\em domain wall fermions}. Shortly thereafter
Narayanan and Neuberger\cite{narayanan_neuberger} provided another realization
of this idea, the {\em overlap} formulation. Interestingly, the relevance of
the wok in ref.~\cite{fu_nielsen} was overlooked, perhaps because it studied
the pure gauge case only. 

What is noteworthy is that, in order to study chiral
lattice fermions in four dimensions, it is necessary to introduce defects that
live in extra dimensions. This is the first time that, in a field-theoretic
context, the need for extra dimensions is required on specific physical
grounds, rather than admitted {\em ad hoc}.
 It is thus interesting to ask the question, 
whether one should consider these extra dimensions as a computational
``trick'' only, or consider
their implications for physics beyond the standard model. I will try to argue
for the latter position. This is particularly relevant, since current
computers and algorithms are starting to come to grips with various systematic
effects of locality (cf., for example, \cite{boyle05})
 so it is useful to consider sources of systematic errors more closely. 

Indeed, in ref.~\cite{KANP} we found that, when the gauge fields become
dynamical, the chiral zero mode disappears in the layered phase. This means
that one cannot ignore the value of the gauge coupling in the extra
dimensions, but must take care to choose it, in conjunction with the value
along the layer, so as to be at a transition surface. Else, the calculated
quantities, as usual, are subject to systematic lattice artifacts: if one
chooses to set the lattice coupling along the extra dimension to zero, one is
in the layered phase and thus the lattice propagator {\em must} be sensitive
to lattice artifacts. If one chooses to take the two couplings equal, one is in
either the bulk Coulomb phase or the bulk confining phase and, far from the
transition, is also sensitive to lattice artifacts. In addition, the
transition of the isotropic theory is first order, since the five-dimensional
isotropic theory is non-renormalizable by power counting. The lattice
artifacts are here unavoidable and large. 
In ref.~\cite{HKAN}  we found numerical evidence 
that the layered to bulk phase transitions was continuous, thus that it is,
indeed, possible to simulate theories with chiral fermions and obtain a
scaling limit. In this limit the coupling constant on the layer depends
parametrically on the coupling constant along the extra dimension(s)-a
scenario reminiscent of the proposals in ref.~\cite{dienes_etal}.

One aspect of domain wall/overlap fermions that has receded in the background
(no pun intended) is the precise nature of the defect that gives rise to the
mass variation along the extra dimensions. It might be useful to look more
closely at specific examples. The simplest case would be that of an additional
scalar field, interacting through Yukawa couplings with the fermions. The
classical solution of the equation of motion for the scalar--in the absence of
the fermions-would be the domain wall, which is then considered as the
background for the fermions. This solution carries a natural topological
charge, related to the chirality of the fermionic zero modes, that live on the
layer. It would be interesting to study the fluctuations of this charge, as
expressed by its susceptibility, for example, by taking into account the
coupling of the scalar field with the fermions, along the lines, for instance,
of ref.\cite{tilak}. This brings us naturally to consider scalar fields
interacting with anisotropic gauge fields and this has, indeed, been done in
the context of the Abelian Higgs model\cite{AAH}, where we mapped the phase
diagram and used the susceptibility to find a continuous phase transition
between the bulk and layered Higgs phases. In ref.~\cite{jansen} the phase
diagram of the Yukawa model with domain wall fermions has been mapped and 
a natural next step would thus be to consider ``gauged'' Yukawa models, 
incorporating both scalar fields and fermions coupled to 
anisotropic gauge fields. 

It has been tempting to use the extra components of the gauge field as
substitutes for the Higgs field. This stems from the fact that, in continuum
language, the term 
$$
-\frac{1}{4g'^2}F_{\mu 5}^2
$$
in the action is the kinetic term of a four-dimensional scalar field, suitably 
rewritten\cite{knechtli}. I would like to argue here that appearances are
misleading and that in the continuum the extra components of the gauge field
decouple. The reason is the following:  
First of all, much of the analysis is carried out using the same value of the
gauge coupling. As indicated above, this cannot be the whole story, since the
isotropic theory is cutoff-dependent. Next, the r\^ole of the extra components
of the gauge field is to trigger confinement through the area law of the
Wilson loops. The lattice formulation makes this quite explicit and shows
that, in the continuum limit these components decouple as fields. They survive
only through the parametric dependence of the four-dimnsional coupling(s) on
the extra-dimensional ones. It is noteworthy that the confinement mechanism of
the chiral zero modes through scalar fields is through the domain walls these
generate--the extra components of the gauge fields don't (and can't) do this. 

Lastly it is necessary to stress that the layered phase exists only for
compact abelian gauge fields, since only they possess both a Coulomb and a
confining phase. Pure Yang-Mills theories with a simple gauge group 
are always confining in all dimensions and thus cannot generate a layered
phase. This means that attempts to use domain wall or overlap fermions coupled
to $SU(2)$ or $SU(3)$ gauge fields, for instance, are subject to systematic
lattice artifacts. These were, until now, smaller than the other systematic
errors, but this is changing\cite{boyle05}. I therefore propose that the
correct way to proceed is to use $U(N)$ instead of  $SU(N)$ lattice gauge
fields. The reason is that $U(N)=U(1)\times SU(N)$. It is the $U(1)$ factor
that will generate the layered phase and localize the fields on the layer. It
is reassuring that the electroweak sector has exactly this structure, for $N=2$!

In conclusion the layered phase of gauge theories coupled to matter fields
provides, on the one hand, a solution to the technical problem of chiral
lattice fermions, on the other hand provides incentive to study the effects of
extra dimensions beyond the classical equations of motion currently used, for
instance,  in brane-world models. 
Indeed it {\em predicts} that brane world models, whose
parameters correspond to bulk phases will be unstable. It sets this question
within reach of concrete numerical simulations. Another open question is the
precise nature of the theory along the transition lines between the bulk and
layered phase(s). Recent numerical simulations\cite{farakos} confirm the
second order nature of the transition between the layered and the bulk Coulomb
phase, so a renormalization group analysis is needed in its vicinity in order
to clarify the field content and obtain concrete numbers.
 Might it be related to 
``little string theories''\cite{little_string_theories}?

\end{document}